# Fluctuations, Trajectory Entropy, and Ziegler's Maximum Entropy Production

V.D. Seleznev, L.M. Martyushev

Institute of Industrial Ecology, Russian Academy of Sciences, 20A  S. Kovalevskaya St., 620219 Ekaterinburg, RUSSIA,
Ural Federal University, 19 Mira St., Ekaterinburg, 620002, RUSSIA,
LeonidMartyushev@gmail.com

We consider relaxation of an isolated system to the equilibrium using detailed balance condition and Onsager's fluctuation approximation. There is a small deviation from the equilibrium in two parameters. For this system, explicit expressions both for the dependence of trajectory entropy on random thermodynamic fluxes and for the dependence of entropy production on the most probable thermodynamic fluxes are obtained. Onsager's linear relations are obtained for the considered model using two methods (maximization of trajectory entropy and Ziegler's maximization of entropy production). Two existing interpretations of the maximum entropy production principle - as a physical principle and as an effective inference procedure - are discussed in the paper.

## 1 Introduction

There are quite a lot of recent examples of successful application of the maximum entropy production principle (MEPP) in physics (kinetic theory of gases, hydrodynamics, theories of crystallization and radiation etc.), biology, and chemistry. Some results of similar investigations can be found, for instance, in reviews [1-3] and the present book. The interest in theoretical grounding of MEPP and in connecting MEPP with other principles [1-12] has become a natural consequence of the achieved success.

MEPP has been considered as a natural generalization (expansion) of the second law of thermodynamics starting from Kohler's and Ziman's papers [13, 14] that used maximization of entropy production for solving the Boltzmann equation. Indeed, if the second law stated the entropy increase in an isolated system for nonequilibrium processes, MEPP stated that this increase would occur to the maximum possible extent. Furthermore, whereas the second law enabled to obtain the basic thermodynamics relations and describe phase transitions in the case of equilibrium (quasistatic) processes, MEPP led to the basic laws of nonequilibrium thermodynamics and enabled to describe nonequilibrium (kinetic) phase transitions [15-17]. Thus, on the basis of the foregoing, MEPP was considered as an independent postulate on entropy substantially complementing and generalizing the second law of thermodynamics. Such interpretation was supported particularly by H. Ziegler, who independently developed one of the MEPP wordings and specified an effective procedure of entropy production variation in the case of fixed thermodynamic forces, which allowed him to obtain the explicit form of thermodynamic fluxes (and particularly, Onsager's linear relations [1,15,16]). Thus, there is a large group of researchers who consider MEPP to be a new and important principle of the physics of nonequilibrium processes; furthermore, a number of scientists (for example, M. Kohler, J. Ziman) proceeded from statistic (kinetic) considerations and the other scientists (for example, H. Ziegler) proceeded from thermodynamic considerations.

However, there is another, a little different view on MEPP. The macroscopic state entropy is associated with the number of microscopic states in the phase space of coordinates and momenta of molecules and atoms satisfying such macroscopic state. In the equilibrium, the number of these microscopic states turns out to be a maximum, and correspondingly the entropy has a maximum value. For certain models, these studies represent a statistical justification of the second law of thermodynamics. Due to C. Shannon and E. Jaynes researches, this kind of view on entropy was generalized: by analogy with the Boltzmann-Gibbs entropy, they introduced the so-called informational entropy applicable to the description of objects of any nature [18,19]. Maximization of the informational entropy enabled to determine the probability of a particular state of the system. Within this trend, the papers (see e.g. [4-11], [20-25]) develop an idea about the relation between the probability of a nonequilibrium process and the number of microscopic trajectories implementing it, about the introduction of the informational entropy as a measure of the number of such microscopic trajectories with the subsequent maximization of that entropy for defining the most probable way of the nonequilibrium system evolution. As consequence, the following opinion was formed: the maximization of the informational entropy written in the phase space of microscopic states for equilibrium is similar to the Boltzmann-Gibbs maximization, and the maximization of the informational entropy in the space of microscopic trajectories for a nonequilibrium process should apparently lead to MEPP. This area of investigations has not reached the required level of rigor so far, though it is very promising for understanding and illustration (using specific models) of the microscopic properties of MEPP and its relations with the other statements existing in the non-equilibrium physics. However, studies in this area sometimes give rise to an idea that MEPP is more or less a simple consequence of the Jaynes information entropy maximization and that MEPP is not an important physical law/principle, but only "an inference algorithm that translates physical assumptions into macroscopic predictions"[1] [4-11]. We will refer to such interpretation of MEPP as an informational interpretation for brevity, as opposed to the above statistical and thermodynamic one.

Thus, there are two opinions about MEPP. In order to obtain a better understanding of these two approaches and be able to compare them it is desirable to find a problem whose solving would enable to independently apply both approaches within the same approximations (formalism): for instance, Ziegler's maximization formalism and an approach through the trajectory entropy maximization. The first approach is purely thermodynamical and the second one is informational (i.e. it is applicable to any scale of description, particularly to the mesoscopic one). Therefore, such investigation is basically possible; however, it has not yet been conducted. The objective of this study is to consider an elementary transfer problem using two methods and to determine the relations and differences between the two approaches maximizing either the trajectory entropy or the entropy production based on Ziegler's procedure.

---

[1] The criticism of such ideas will be set forth in the conclusion hereof.

The paper consists of four parts. The second part introduces a model, its basic assumptions, and the fundamental equations obtained within the scope thereof. This part has primarily a methodological purpose. The third and fourth parts consider the problem through the trajectory entropy maximization method and using Ziegler's procedure, respectively. This is the central section of the paper achieving the objective set forth herein. The conclusion of the article contains, in addition to the main results, a brief reasoning on the various approaches to the interpretation of MEPP.

## 2 Onzager's model and linear thermodynamic relations

Let us consider an isolated system with possible fluctuations. An arbitrary (nonequilibrium) macroscopic state of the system will be described by the set of parameters $A_i(t)$ that acquire the values $A_i^{eq}$ in the equilibrium state. Let us designate the difference between the parameter values and their equilibrium values as $\alpha_i = A_i(t) - A_i^{eq}$. In the nonequilibrium case, $\alpha_i \neq 0$. For brevity of further description, it would be convenient to introduce the vector $\boldsymbol{\alpha}$ with the macroscopic parameters $\alpha_i$ serving as its components. Herein we shall assume that the observation time $\tau$ is much smaller than the time of the system relaxation to the equilibrium, i.e. we shall consider the "momentary" response of the system to the nonequilibrium state generated therein.

Any state of the system with $\boldsymbol{\alpha}$ can be characterized by the number of microscopic states $\Gamma(\boldsymbol{\alpha})$, which, as is known [26,27], is a maximum in the equilibrium. As usual, let us assume that all microscopic states corresponding to that macroscopic state are equiprobable. Let us define the probability of finding the equilibrium system in the state with $\boldsymbol{\alpha}$ as:

$$W(\boldsymbol{\alpha}) \propto \Gamma(\boldsymbol{\alpha}) \tag{1}$$

The process of the system relaxation to the equilibrium will be characterized by the conditional probability that the system being in the state $\boldsymbol{\alpha_0}$ at the initial moment of time will get to the state $\boldsymbol{\alpha}$ at the moment of time $\tau$ [27]. Let us designate this probability as $P(\boldsymbol{\alpha_0}|\boldsymbol{\alpha},\tau)$ (Fig.1). As is in the previous case, we shall assume that this conditional (trajectory) probability is proportional to the number of microscopic trajectories of the transition from the state $\boldsymbol{\alpha_0}$ to the state $\boldsymbol{\alpha}$ for the time $\tau$, i.e.

$$P(\boldsymbol{\alpha_0}|\boldsymbol{\alpha},\tau) \propto \Gamma(\boldsymbol{\alpha_0}|\boldsymbol{\alpha},\tau). \qquad (2)$$

Let the transition from $\boldsymbol{\alpha_0}$ to $\boldsymbol{\alpha}^*(\boldsymbol{\alpha_0})$ for the time $\tau$ be the most probable (according to (2), it means that such a transition occurs with the maximum number of microscopic trajectories).

Let us use the classical method [26] to determine the entropy in the state $\boldsymbol{\alpha}$ and in the equilibrium ($\boldsymbol{\alpha}=\boldsymbol{0}$): $S(\boldsymbol{\alpha})=\ln\Gamma(\boldsymbol{\alpha})$ and $S(0)=\ln\Gamma(0)$, then:

$$\Delta S(\alpha) = S(\alpha) - S(0) = \ln\frac{\Gamma(\alpha)}{\Gamma(0)} = \ln\frac{W(\alpha)}{W(0)} \qquad (3)$$

By analogy, let us introduce the trajectory entropies $S_{tr}(\boldsymbol{\alpha_0}|\boldsymbol{\alpha},\tau) = \ln\Gamma(\boldsymbol{\alpha_0}|\boldsymbol{\alpha},\tau)$ and $S_{tr}(\boldsymbol{\alpha_0}|\boldsymbol{\alpha}^*(\alpha_0),\tau) = \ln\Gamma(\boldsymbol{\alpha_0}|\boldsymbol{\alpha}^*(\alpha_0),\tau)$, then:

$$\begin{aligned}\Delta S_{tr}(\boldsymbol{\alpha_0}|\boldsymbol{\alpha},\tau) &= S_{tr}(\boldsymbol{\alpha_0}|\boldsymbol{\alpha},\tau) - S_{tr}(\boldsymbol{\alpha_0}|\boldsymbol{\alpha}^*(\alpha_0),\tau) = \\ &= \ln\frac{\Gamma(\boldsymbol{\alpha_0}|\boldsymbol{\alpha},\tau)}{\Gamma(\boldsymbol{\alpha_0}|\boldsymbol{\alpha}^*(\alpha_0),\tau)} = \ln\frac{P(\boldsymbol{\alpha_0}|\boldsymbol{\alpha},\tau)}{P(\boldsymbol{\alpha_0}|\boldsymbol{\alpha}^*(\alpha_0),\tau)}\end{aligned} \qquad (4)$$

As a result:

$$P(\boldsymbol{\alpha_0}|\boldsymbol{\alpha},\tau) = P(\boldsymbol{\alpha_0}|\boldsymbol{\alpha}^*(\alpha_0),\tau) \cdot e^{\Delta S_{tr}(\boldsymbol{\alpha_0}|\boldsymbol{\alpha},\tau)} \qquad (5)$$

Processes for which $P(\boldsymbol{\alpha_0}|\boldsymbol{\alpha}^*(\alpha_0),\tau)$ is a constant independent of $\boldsymbol{\alpha_0}$ will be considered below. Such approximation is quite common [27, 28].

In the equilibrium, the number of system transitions in the forward $\boldsymbol{\alpha_0} \to \boldsymbol{\alpha}$ and reverse $\boldsymbol{\alpha} \to \boldsymbol{\alpha_0}$ directions during the macroscopic time $\tau$ should be equal, correspondingly it can be shown [27,29] that the so-called principle of detailed balance (equilibrium) relating the probability of finding the system in a certain macroscopic state to the conditional probability of a transition therefrom is true:

$$W(\boldsymbol{\alpha_0})P(\boldsymbol{\alpha_0}|\boldsymbol{\alpha},\tau) = W(\boldsymbol{\alpha})P(\boldsymbol{\alpha}|\boldsymbol{\alpha_0},\tau). \qquad (6)$$

Condition (6) is proved for fluctuations of the equilibrium system [27,29]. Following the classical studies of L. Onsager [29] and the monograph [27] where this approach is stated in the most complete form, let us suppose (Onsager's hypothesis) that this relation is also true for *near* equilibrium conditions. In other words, the evolution of the fluctuating equilibrium system found itself in the state $\boldsymbol{\alpha_0}$ will be similar to the evolution of the specifically prepared (close to equilibrium) system brought to the same state $\boldsymbol{\alpha_0}$, and then left for spontaneous relaxation.

Further, in order to avoid cumbersome calculations, let us consider that $\boldsymbol{\alpha}$ has only two components. It should be noted that the calculations below can be generalized for any number of components. According to (6), we have:

$$W(\alpha_1,\alpha_2)/W(\alpha_{10},\alpha_{20}) = P(\alpha_{10},\alpha_{20}|\alpha_1,\alpha_2,\tau)/P(\alpha_1,\alpha_2|\alpha_{10},\alpha_{20},\tau). \qquad (7)$$

Using (3)-(5), Eq. (7) can also be written in the form:

$$\Delta S(\boldsymbol{\alpha}) - \Delta S(\boldsymbol{\alpha_0}) = \Delta S_{tr}(\boldsymbol{\alpha_0}|\boldsymbol{\alpha},\tau) - \Delta S_{tr}(\boldsymbol{\alpha}|\boldsymbol{\alpha_0},\tau). \qquad (8)$$

In the case of L. Onsager's classical treatment [29], the principle of detailed balance together with the supposition of a linear relation between the mean *change* of $\boldsymbol{\alpha}$ for the time $\tau$ and the quantity $\boldsymbol{\alpha}$ itself leads to the proof of the so-called reciprocal relations and to the Gaussian form of $W(\boldsymbol{\alpha})$ and $P(\boldsymbol{\alpha_0}|\boldsymbol{\alpha},\tau)$. Here we consider an inverse problem: the Gaussian form of $W(\boldsymbol{\alpha})$ and $P(\boldsymbol{\alpha_0}|\boldsymbol{\alpha},\tau)$ is postulated and the linear relations for the change of $\boldsymbol{\alpha}$ are obtained using the principle of detailed balance. Such a statement of the problem does not pretend to essential originality (in fact, such possibility seems quite obvious); however, in our opinion, such treatment is the simplest and the shortest way to explicitly express the trajectory entropy, the entropy production, as well as other required quantities using the model parameters. It enables, in the easiest way, to achieve the objective set forth in the introduction (see sections 3 and 4).

So, minor deviations from the equilibrium will be assumed. In this case, the so-called Gauss distribution is a frequently used approximation for the equilibrium deviation probability [26, 27]:

$$W(\boldsymbol{\alpha}) = \Omega \exp(-\beta_{11}\alpha_1^2 - \beta_{22}\alpha_2^2 - 2\beta_{12}\alpha_1\alpha_2), \qquad (9)$$

and for the trajectory probability [27-29]:

$$P(\boldsymbol{\alpha_0}|\boldsymbol{\alpha},\tau) = \Xi \cdot \exp\{-\gamma_{11}(\alpha_1^*(\boldsymbol{\alpha_0},\tau)-\alpha_1)^2 \\ -\gamma_{22}(\alpha_2^*(\boldsymbol{\alpha_0},\tau)-\alpha_2)^2 - 2\gamma_{12}(\alpha_1^*(\boldsymbol{\alpha_0},\tau)-\alpha_1)(\alpha_2^*(\boldsymbol{\alpha_0},\tau)-\alpha_2)\}, \quad (10)$$

where $\Omega$, $\Xi$ are normalization constants (independent of $\boldsymbol{\alpha},\boldsymbol{\alpha_0}$, but dependent, particularly, on $A_i^{eq}$); $\beta_{ij}$ is a coefficient that is inversely proportional to the distribution variance of the random quantity $\boldsymbol{\alpha}$ relative to the equilibrium value ($\boldsymbol{\alpha}=0$); $\gamma_{ij}$ is a coefficient[2] that is inversely proportional to the distribution variance relative to the average (most probable) value $\alpha_i^*(\boldsymbol{\alpha_0},\tau)$ during the transition from the point $\boldsymbol{\alpha_0}$ for the time $\tau$. It should be noted that the variances in this approximation are assumed independent of $\boldsymbol{\alpha_0}$, $\boldsymbol{\alpha}$ [27-29]. It should be also emphasized that $\beta_{ij}$ does not dependent on the time $\tau$ (as it characterizes the fluctuation in the equilibrium state); $\gamma_{ij}$, on the contrary, depends on the time and, moreover, substantially increases with the decrease of $\tau$ (for $\tau \to 0$, $\alpha_i^* \to \alpha_{i0}$ and the distribution tends to a delta function). Let us accept a simple supposition that $\gamma_{ij} = \gamma_{ij}^0/\tau$ [27-29], where $\gamma_{ij}^0$ is some constant.

By inserting (9) and (10) into (7), we will obtain:

$$\frac{\exp(-\beta_{11}\alpha_1^2 - \beta_{22}\alpha_2^2 - 2\beta_{12}\alpha_1\alpha_2)}{\exp(-\beta_{11}\alpha_{10}^2 - \beta_{22}\alpha_{20}^2 - 2\beta_{12}\alpha_{10}\alpha_{20})} = \\ = \frac{\exp(-\gamma_{11}(\alpha_1^*(\boldsymbol{\alpha_0},\tau)-\alpha_1)^2 - \gamma_{22}(\alpha_2^*(\boldsymbol{\alpha_0},\tau)-\alpha_2)^2 - 2\gamma_{12}(\alpha_1^*(\boldsymbol{\alpha_0},\tau)-\alpha_1)(\alpha_2^*(\boldsymbol{\alpha_0},\tau)-\alpha_2))}{\exp(-\gamma_{11}(\alpha_1^*(\boldsymbol{\alpha},\tau)-\alpha_{10})^2 - \gamma_{22}(\alpha_2^*(\boldsymbol{\alpha},\tau)-\alpha_{20})^2 - 2\gamma_{12}(\alpha_1^*(\boldsymbol{\alpha},\tau)-\alpha_{10})(\alpha_2^*(\boldsymbol{\alpha},\tau)-\alpha_{20}))}. \quad (11)$$

Here $\alpha_i^*(\boldsymbol{\alpha},\tau)$ is the most probable value in the case of the transition from $\boldsymbol{\alpha}$ during the time $\tau$.

By finding the logarithm of the latter, we will obtain:

$$\beta_{11}(\alpha_{10}^2 - \alpha_1^2) + \beta_{22}(\alpha_{20}^2 - \alpha_2^2) + 2\beta_{12}(\alpha_{10}\alpha_{20} - \alpha_1\alpha_2) = \\ = \gamma_{11}((\alpha_1^*(\boldsymbol{\alpha},\tau)-\alpha_{10})^2 - (\alpha_1^*(\boldsymbol{\alpha_0},\tau)-\alpha_1)^2) + \gamma_{22}((\alpha_2^*(\boldsymbol{\alpha},\tau)-\alpha_{20})^2 - (\alpha_2^*(\boldsymbol{\alpha_0},\tau)-\alpha_2)^2) + \\ + 2\gamma_{12}((\alpha_1^*(\boldsymbol{\alpha},\tau)-\alpha_{10})(\alpha_2^*(\boldsymbol{\alpha},\tau)-\alpha_{20}) - (\alpha_1^*(\boldsymbol{\alpha_0},\tau)-\alpha_1)(\alpha_2^*(\boldsymbol{\alpha_0},\tau)-\alpha_2)). \quad (12)$$

---

[2] According to their definition [27-29]: $\gamma_{ii} > 0$ and $\gamma_{11}\gamma_{22} - \gamma_{12}^2 \geq 0$. It meets the requirement of a non-negative power exponent for the two-dimensional Gauss (normal) distribution.

If the definitions of the state entropy and the trajectory entropy introduced above, as well as Eq. (8) are recalled, then the taking of logarithm of Eq.(11) allows obtaining:

$$\Delta S(\boldsymbol{\alpha_0}) = -\beta_{11}\alpha_{10}^2 - \beta_{22}\alpha_{20}^2 - 2\beta_{12}\alpha_{10}\alpha_{20}, \tag{13}$$

$$\Delta S(\boldsymbol{\alpha}) = -\beta_{11}\alpha_1^2 - \beta_{22}\alpha_2^2 - 2\beta_{12}\alpha_1\alpha_2, \tag{14}$$

$$\Delta S_{tr}(\boldsymbol{\alpha_0}|\boldsymbol{\alpha},\tau) = -\gamma_{11}(\alpha_1^*(\boldsymbol{\alpha_0},\tau)-\alpha_1)^2 - \gamma_{22}(\alpha_2^*(\boldsymbol{\alpha_0},\tau)-\alpha_2)^2, \tag{15}$$
$$-2\gamma_{12}(\alpha_1^*(\boldsymbol{\alpha_0},\tau)-\alpha_1)(\alpha_2^*(\boldsymbol{\alpha_0},\tau)-\alpha_2))$$

$$\Delta S_{tr}(\boldsymbol{\alpha}|\boldsymbol{\alpha_0},\tau) = -\gamma_{11}(\alpha_1^*(\boldsymbol{\alpha},\tau)-\alpha_{10})^2 - \gamma_{22}(\alpha_2^*(\boldsymbol{\alpha},\tau)-\alpha_{20})^2. \tag{16}$$
$$-2\gamma_{12}(\alpha_1^*(\boldsymbol{\alpha},\tau)-\alpha_{10})(\alpha_2^*(\boldsymbol{\alpha},\tau)-\alpha_{20}))$$

In order to find the relationship between the forward and reverse trajectories, let us transform (12); for this purpose, we will consider very small times $\tau$ and apply the Taylor expansion[3]:

$$\alpha_i^*(\boldsymbol{\alpha_0},\tau) = \alpha_{i0} + \tau \left.\frac{\partial \alpha_i^*(\boldsymbol{\alpha_0},\tau)}{\partial \tau}\right|_{\tau=0} + \ldots$$

We will apply two consecutive expansions for the quantity below: first, in terms of $\tau$ near zero, then in terms of $\alpha_i$ near $\alpha_{i0}$:

$$\alpha_i^*(\boldsymbol{\alpha},\tau) = \alpha_i + \tau \left.\frac{\partial \alpha_i^*(\boldsymbol{\alpha},\tau)}{\partial \tau}\right|_{\tau=0} + \ldots =$$
$$= \alpha_i + \tau \left.\frac{\partial \alpha_i^*(\boldsymbol{\alpha_0},\tau)}{\partial \tau}\right|_{\tau=0} + \tau \sum_{i=1}^{2} \left.\frac{\partial \alpha_i^*(\boldsymbol{\alpha},\tau)}{\partial \tau \partial \alpha_i}\right|_{\tau=0,\alpha_i=\alpha_{i0}} (\alpha_i - \alpha_{i0}) + \ldots$$

---

[3] The following uses the fact that the initially specified value is the most probable value for the deviation from the equilibrium at the initial moment of time $\alpha_i^*(\boldsymbol{\alpha_0},0) = \alpha_{i0}$.

By neglecting the second orders of smallness ($\propto \tau(\alpha_i - \alpha_{i0})$), we will obtain:

$$\alpha_i^*(\boldsymbol{a_0}, \tau) - \alpha_{i0} = \alpha_i^*(\boldsymbol{a}, \tau) - \alpha_i. \quad (17)$$

Let us introduce the following notations: $\Delta\alpha_i^* = \alpha_{i0} - \alpha_i^*(\boldsymbol{a_0}, \tau); \Delta\alpha_i = \alpha_{i0} - \alpha_i$. These quantities characterize the most probable and actual (random) change of values of the parameters $A_i(\tau)$ relative to the original $A_i(0)$ during the time $\tau$. By using them jointly with (17), we will obtain:

$$\alpha_i^*(\boldsymbol{a_0}, \tau) - \alpha_i = \Delta\alpha_i - \Delta\alpha_i^*; \quad (18)$$
$$\alpha_i^*(\boldsymbol{a}, \tau) - \alpha_{i0} = (\alpha_i^*(\boldsymbol{a_0}, \tau) - \alpha_{i0} + \alpha_i) - \alpha_{i0} = -\Delta\alpha_i^* - \Delta\alpha_i$$

Using two last expressions, Eq. (12) can be reduced to the form[4]:

$$2\Delta\alpha_1(\alpha_{10}\beta_{11} + \beta_{12}\alpha_{20}) + 2\Delta\alpha_2(\beta_{22}\alpha_{20} + \beta_{12}\alpha_{10}) = \quad (19)$$
$$= 4\Delta\alpha_1(\gamma_{11}\Delta\alpha_1^* + \gamma_{12}\Delta\alpha_2^*) + 4\Delta\alpha_2(\gamma_{22}\Delta\alpha_2^* + \gamma_{12}\Delta\alpha_1^*).$$

Since the deviations ($\Delta\alpha_1, \Delta\alpha_2$) from the initial state are independent, then the following can be obtained from equality Eq.(19):

$$2(\beta_{11}\alpha_{10} + \beta_{12}\alpha_{20}) = 4\gamma_{11}\Delta\alpha_1^* + 4\gamma_{12}\Delta\alpha_2^* \quad (20)$$
$$2(\beta_{22}\alpha_{20} + \beta_{12}\alpha_{10}) = 4\gamma_{22}\Delta\alpha_2^* + 4\gamma_{12}\Delta\alpha_1^*.$$

Let us introduce a number of important quantities. Since the system is considered as isolated, then its entropy change rate proves to be equal to the entropy production $\Sigma$ [1, 27]. Let us use the classical method [26, 27, 29] to introduce the thermodynamic forces $X_i$ and fluxes $J_i$:

$$\Sigma = \frac{dS(\boldsymbol{a})}{dt} = \frac{\partial S}{\partial \alpha_1}\frac{d\alpha_1}{dt} + \frac{\partial S}{\partial \alpha_2}\frac{d\alpha_2}{dt} = X_1 J_1 + X_2 J_2, \quad (21)$$

---

[4] For small $\tau$ values: $\beta_{ij}\Delta\alpha_i\Delta\alpha_j \ll \gamma_{ij}\Delta\alpha_i\Delta\alpha_j^*$ (because $\gamma_{ij} = \gamma_{ij}^0/\tau$).

where:
$$X_i = -\partial S / \partial \alpha_i \qquad (22)$$
$$J_i = -d\alpha_i / dt \qquad (23)$$

According to (13) and (22), the thermodynamic forces acting in the system at the initial moment of relaxation to the equilibrium, are equal to:

$$X_1 = 2(\beta_{11}\alpha_{10} + \beta_{12}\alpha_{20})$$
$$X_2 = 2(\beta_{22}\alpha_{20} + \beta_{12}\alpha_{10}) \qquad (24)$$

Using (23), (24)[5], and the relation between $\gamma_{ij}$ and $\gamma_{ij}^0$, expression (20) can be rewritten in the form:

$$X_1 = 4\gamma_{11}^0 J_1^* + 4\gamma_{12}^0 J_2^*$$
$$X_2 = 4\gamma_{12}^0 J_1^* + 4\gamma_{22}^0 J_2^*, \qquad (25)$$

or by transforming, we will obtain:

$$J_1^* = L_{11}X_1 + L_{12}X_2$$
$$J_2^* = L_{21}X_1 + L_{22}X_2, \qquad (26)$$

where the following kinetic coefficients are introduced:

$$L_{11} = \gamma_{22}^0 / 4(\gamma_{22}^0\gamma_{11}^0 - (\gamma_{12}^0)^2),$$
$$L_{12} = L_{21} = -\gamma_{12}^0 / 4(\gamma_{22}^0\gamma_{11}^0 - (\gamma_{12}^0)^2), \qquad (27)$$
$$L_{22} = \gamma_{11}^0 / 4(\gamma_{11}^0\gamma_{22}^0 - (\gamma_{12}^0)^2).$$

Based on the properties of $\gamma_{ij}^0$ (see footnote 2): $L_{ij} > 0$ and $L_{11}L_{22} - L_{12}^2 \geq 0$.

Thus, it is shown that the assumption of Gaussianity (9), (10) and the principle of detailed balance (6) result in the known linear Onsager relations [26, 27, 29]. This relation links the most probable flux in the system with the thermodynamic force existing within the time interval $\tau$. The reasoning set forth herein is the study of the problem opposite to the problem considered by Onsager [29].

---

[5] For small τ values: $d\alpha_i^* / dt \approx -\Delta\alpha_i^* / \tau$. The minus sign has resulted from the fact that $\Delta\alpha_i^*$ was introduced as a difference between the initial and final value.

We conclude the present section with a number of useful relations following from the above equations. According to Eqs. (17), (18), let us rewrite Eqs. (15), (16) in the form:

$$\Delta S_{tr}(\boldsymbol{\alpha_0}|\boldsymbol{\alpha},\tau) = -\gamma_{11}(\Delta\alpha_1 - \Delta\alpha_1^*)^2 - \gamma_{22}(\Delta\alpha_2 - \Delta\alpha_2^*)^2 - \qquad (28)$$
$$- 2\gamma_{12}(\Delta\alpha_1 - \Delta\alpha_1^*)(\Delta\alpha_2 - \Delta\alpha_2^*)$$

$$\Delta S_{tr}(\boldsymbol{\alpha}|\boldsymbol{\alpha_0},\tau) = -\gamma_{11}(\Delta\alpha_1 + \Delta\alpha_1^*)^2 - \gamma_{22}(\Delta\alpha_2 + \Delta\alpha_2^*)^2 - \qquad (29)$$
$$- 2\gamma_{12}(\Delta\alpha_1 + \Delta\alpha_1^*)(\Delta\alpha_2 + \Delta\alpha_2^*)$$

According to (8) and (21), the following can be written for small $\tau$ values:

$$\tau\Sigma = S(\boldsymbol{\alpha}) - S(\boldsymbol{\alpha_0}) = \Delta S_{tr}(\boldsymbol{\alpha_0}|\boldsymbol{\alpha},\tau) - \Delta S_{tr}(\boldsymbol{\alpha}|\boldsymbol{\alpha_0},\tau) \qquad (30)$$

or according to (21)-(24):

$$\tau\Sigma = X_1\Delta\alpha_1 + X_2\Delta\alpha_2 = \qquad (31)$$
$$= 2\Delta\alpha_1(\alpha_{10}\beta_{11} + \beta_{12}\alpha_{20}) + 2\Delta\alpha_2(\beta_{22}\alpha_{20} + \beta_{12}\alpha_{10})$$

Now let us proceed to extreme properties of the trajectory entropy and the entropy production for the model under consideration.

**3 Informational approach: trajectory entropy maximization**

When describing nonequilibrium processes using the informational approach, first some entropy (for example, trajectory entropy) is extremized in order to define the trajectories distribution function in the phase space. Such a procedure is performed subject to the existing (or supposed) constraints (relations between the quantities more or less evident for the process under study). Then the found distribution function is used for calculation of the needed nonequilibrium properties of the process. This classical procedure was repeatedly described in the literature (see, e.g. [4]). The unfalsifiability (according to K. Popper) of that procedure is its fundamental disadvantage [7,8,10]. Indeed, if the calculation results for nonequilibrium properties fail to match the experimental data, it is considered that the constraints used for maximization were incorrectly selected. These constraints (relationships) are adjusted and the procedure is repeated. As a result, the informational approach appears to be some kind of method aimed at selecting the result not contradicting the known experiments. Obviously, the unfalsifiability could be a serious disadvantage of the mentioned approach (casting serious doubt on the scientific nature of the method); however, this method is considered to be only a procedure (mathematical algorithm) for passive translation of physical assumptions (constraints) into predictions without introducing any additional physical assumptions [7]. Thus, such a method is some kind of mathematical device. However, this mathematical procedure entails problems of purely mathematical nature. The first problem of such a procedure is its mathematical unreasonableness. Thus, the "liberty of action" (when selecting constraints) may either yield no desired results at all or allow several solutions that meet the selected criteria but substantially differ in terms of both used constraints and predictions (outside the scope of the selected criteria). The second problem is connected with the choice of the measure of information and, correspondingly, the formula of informational entropy. From the logical viewpoint, there is no best option. There are multiple variants besides the Shannon formula; and many of them prove to be useful in different applications [30].

Despite the mentioned fundamental disadvantages, we will maximize the trajectory entropy for the problem under consideration. This is justified by the fact that the studied model is fairly simple and, as consequence, the existing constraints are accurately set.

In the model at hand, the trajectory entropy as a function of random deviation of $\boldsymbol{\alpha}$ has the form (15) or (28). The explicit form of this entropy was obtained using the Gaussian form of the distribution function $P(\boldsymbol{\alpha_0}|\boldsymbol{\alpha},\tau)$ and the supposition of detailed balance. As consequence, the trajectory entropy depends not on the distribution function but on other variables. In this case, the trajectory entropy maximization should lead to the finding of equations linking these variables. The maximization can only formally be considered as unconstrained because all constraints have already been introduced to the expression of trajectory entropy when obtaining its explicit form.

According to Eq.(4), the trajectory entropy for the forward trajectory $S_{tr}(\boldsymbol{\alpha_0}|\boldsymbol{\alpha},\tau)$ is related to $\Delta S_{tr}(\boldsymbol{\alpha_0}|\boldsymbol{\alpha},\tau)$ and to the trajectory entropy for the most probable forward trajectory $S_{tr}^*(\boldsymbol{\alpha_0}|\boldsymbol{\alpha}^*(\boldsymbol{\alpha_0}),\tau)$ as:

$$\frac{\partial S_{tr}(\boldsymbol{\alpha_0}|\boldsymbol{\alpha},\tau)}{\partial \Delta\alpha_i} = \frac{\partial (S_{tr}^*(\boldsymbol{\alpha_0}|\boldsymbol{\alpha}^*(\boldsymbol{\alpha}_0),\tau) + \Delta S_{tr}(\boldsymbol{\alpha_0}|\boldsymbol{\alpha},\tau))}{\partial \Delta\alpha_i} = \frac{\partial \Delta S_{tr}(\boldsymbol{\alpha_0}|\boldsymbol{\alpha},\tau)}{\partial \Delta\alpha_i} \quad (32)$$

By inserting the explicit form $\Delta S_{tr}(\boldsymbol{\alpha_0}|\boldsymbol{\alpha},\tau)$ (Eq. (28)), we will set (32) equal to zero. It can easily be obtained that the trajectory entropy maximum conforms to $\Delta\alpha_i = \Delta\alpha_i^*$. According to (10) and (18), the most probable process trajectory also conforms to the condition $\Delta\alpha_i = \Delta\alpha_i^*$.

Since the maximum of the trajectory entropy deviation is obtained in the case $\Delta\alpha_i = \Delta\alpha_i^*$, then using (29)-(31) it is easy to obtain:

$$\left.\frac{\partial \Delta S_{tr}(\boldsymbol{a_0}|\boldsymbol{a},\tau)}{\partial \Delta \alpha_i}\right|_{\Delta \alpha_i = \Delta \alpha_i^*} = \left.\frac{\partial (\tau \Sigma + \Delta S_{tr}(\boldsymbol{a}|\boldsymbol{a_0},\tau))}{\partial \Delta \alpha_i}\right|_{\Delta \alpha_i = \Delta \alpha_i^*} =$$

$$= \left.\frac{\partial (X_i \Delta \alpha_i + \Delta S_{tr}(\boldsymbol{a}|\boldsymbol{a_0},\tau))}{\partial \Delta \alpha_i}\right|_{\Delta \alpha_i = \Delta \alpha_i^*} = X_i - 4\gamma_{ii}\Delta \alpha_i^* - 4\gamma_{ij}\Delta \alpha_j^* = 0,$$

$$X_i = 4\gamma_{ii}\Delta \alpha_i^* + 4\gamma_{ij}\Delta \alpha_j^*. \tag{33}$$

This expression coincides with Eq.(25). Thus, Onsager's linear relations correspond to the trajectory entropy maximum.

The results obtained from the maximization can be interpreted in two ways. On the one hand, they indicate internal consistency of the used informational method because any other obtained result different from (33) could be deemed, at best, the consequence of false transformations and, at worst, another (in this case, logical) disadvantage of the informational approach[6]. On the other hand, the obtained result certainly indicates that for the model under study the trajectory entropy maximization with a number of constraints allows finding the most probable macrotrajectory satisfying the valid law of relation between thermodynamic fluxes and forces (33). This points to the possibility of generalizing the conventional method of equilibrium entropy maximization.

**4 Thermodynamic approach: Ziegler's maximization**

As opposed to the above method, this approach focuses immediately on the search for relationships between the most probable quantities. Random quantities and their distribution functions are beyond the scope of the approach. The entropy production is considered to be a known function of thermodynamic fluxes [1, 15,16]. The relationship between the thermodynamic fluxes and forces is searched through the maximization of entropy production in the space of independent fluxes at the fixed thermodynamic forces [1, 15,16]. As opposed to the informational approach, Ziegler's method is falsifiable. Indeed, the entropy production is a well-defined macroscopic property of the system connected with the energy dissipation to heat. Thermodynamic forces and fluxes also have a clear physical meaning and are measurable in the experiment. As consequence, if the entropy production maximization based on Ziegler's method yields predictions different from the experiment (particularly, relationships between the fluxes and forces), then it will be possible to disprove the principle (or at least limit the scope of its validity). The fact that for specific systems the finding of the entropy production as a function of fluxes is not always an easy task can be reckoned among the disadvantages of this thermodynamic principle: there are no standard and formal procedures, each specific case requires an individual approach. The fact that H. Ziegler has developed his method only for the systems with biunique correspondence between the thermodynamic force and flux represents another disadvantage.

---

[6] Indeed, as a logical consequence, as it is shown in section 2, the considered model contains the linear relations of forces and fluxes, as well as the form of the trajectory entropy. The disagreement between the result obtained within the trajectory entropy maximization and (25) would therefore represent a serious problem.

Formally, Eqs. (25) or (26) were many times obtained using Ziegler's procedure [1, 15,16]. For this purpose, the entropy production as a bilinear function of thermodynamic fluxes was *postulated* and then the maximization was carried out with the fixed forces. In the present study, it is possible, within the scope of the model under consideration, to *explicitly obtain* both the form of the entropy production and the constraint for the maximization. We will show it. Let us rewrite the detailed balance relation (30) for the most probable trajectory $(\Delta \alpha_i = \Delta \alpha_i^*)$. In this case, $\Delta S_{tr}(\boldsymbol{\alpha_0}|\boldsymbol{\alpha}, \tau) = 0$ (see Eq.(28)), and as a result (see Eq.(29)):

$$\tau \Sigma = -\Delta S_{tr}(\boldsymbol{\alpha}|\boldsymbol{\alpha_0}, \tau) = 4\gamma_{11}\Delta \alpha_1^{*2} + 4\gamma_{22}\Delta \alpha_2^{*2} + 8\gamma_{12}\Delta \alpha_1^* \Delta \alpha_2^*$$

or using Eqs. (23), (31) we will obtain:

$$X_1 \Delta \alpha_1^* + X_2 \Delta \alpha_2^* = 4\gamma_{11}\Delta \alpha_1^{*2} + 4\gamma_{22}\Delta \alpha_2^{*2} + 8\gamma_{12}\Delta \alpha_1^* \Delta \alpha_2^*,$$

$$X_1 J_1^* + X_2 J_2^* = 4\gamma_{11}^0 J_1^{*2} + 4\gamma_{22}^0 J_2^{*2} + 8\gamma_{12}^0 J_1^* J_2^*. \tag{34}$$

Obviously, the left-hand side of the last expression contains the entropy production $\Sigma(J^*)$ through thermodynamic forces, whereas the entropy production on the right-hand side is written in the space of thermodynamic fluxes only. It is easy to check that the constrained entropy production maximum $\Sigma(J^*) = 4\gamma_{11}^0 J_1^{*2} + 4\gamma_{22}^0 J_2^{*2} + 8\gamma_{12}^0 J_1^* J_2^*$ subject to (34) (where $\mu$ is the Lagrange multiplier):

$$\frac{\partial}{\partial J_i^*}\left(\Sigma(J^*) - \mu(\Sigma(J^*) - X_1 J_1^* - X_2 J_2^*)\right) = 0, \tag{35}$$

also leads to Onsager's linear relations (25). This procedure is often given in the literature [1, 15,16] and will not be repeated here.

# 5 Conclusion

The present paper considers the simplest model of a slightly nonequilibrium system with two thermodynamic forces that satisfy the detailed balance condition and with the equilibrium and the transition probabilities distributed according to Gauss. Within the scope of this model, we have managed to:

- Explicitly write a formula for the trajectory entropy as a function of random deviations from the equilibrium. For the first time, it enabled to show that the maximization of the entropy of random microtrajectories leads to Onsager's linear relations.

- Explicitly obtain a formula for the entropy production as a function of thermodynamic fluxes, which is a necessary procedure for the entropy production maximization according to Ziegler. The obtained expression agrees with the expression that was previously postulated and used by H. Ziegler for obtaining Onsager's linear relations.

Thus, it is shown that the phenomenological transfer equations (Onsager's linear relations) can be independently obtained using at least two methods: Ziegler's procedure, or the trajectory entropy maximization. Within the scope of the considered problem, it can not be stated that the method using the trajectory entropy maximization is more general or that Ziegler's maximization procedure follows from this method (and the opposite is not true). It can be rather concluded that these are two **different** methods[7] leading to the same result. The model within the scope of which the solutions are given is the only link between the considered methods. The model contains the linear relationship between fluxes and forces as shown in Sect. 2. That is what may suggest excessiveness of the suppositions used in Sections 3 and 4. It is, however, certainly not the case. Here the following analogy from mechanics can be drawn. Newton's laws allow solving any mechanical problem; however, these laws can be used for developing new ways to the solution of mechanical problems using the variational methods (Hamilton, Lagrange, etc.). These new methods will not represent excessive formulations for solution of the original mechanical problem; in a number of cases, they just prove to be more convenient for solution of problems.

In conclusion, let us take this opportunity to express **our view** about MEPP and two versions of MEPP interpretation mentioned in the Introduction.

1) Maximum entropy production principle is an important **physical principle** following from the generalization of experimental data. This is particularly confirmed by multiple examples given in the papers [1-3]. In this context, that principle resembles the second law of thermodynamics, essentially generalizing it. Its possible wording is as follows: *at each level of description, with preset external constraints, the relationship between the cause and the response of a nonequilibrium system is established in order to maximize the entropy production density* [17, 31]. This wording substantially generalizes the formulation previously proposed by H. Ziegler and discussed herein in detail within the scope of the particular model. Specifically, the requirement for biunique correspondence between the thermodynamic fluxes and forces can be reckoned among the disadvantages of Ziegler's formulation. This requirement has substantially limited Ziegler' principle, depriving it of important and interesting fields of application (particularly, study of nonequilibrium phase transitions [17, 31]).

---

[7] For one approximation, extremization is carried out for a random deviation from the equilibrium, whereas for the other approximation, it is carried out for a thermodynamic flux, i.e. the most probable deviation.

2) It will be recalled that the second law of thermodynamics (essentially equivalent to the statement of positivity of entropy production) allows identifying "one-way direction" or "asymmetry" of time [32]. As it is known, TIME is the most complex and versatile physical quantity that still lacks a universally acknowledged definition [32-34]. Despite having learned to measure time, we still fail to understand its nature. In this regard, we have a conviction that the consideration of MEPP as a new and important addition to the second law of thermodynamics[8] will enable to progress in the understanding of new properties of time. Here the papers [12, 31] where the origin of MEPP is associated with the hypothesis of independence of the entropy production sign when transforming the time scale (that influence the reference system of fluxes), can be mentioned.

3) MEPP is a relatively new principle. Therefore, the range of its applicability is not fully understood. The constraints of the principle should be searched for based primarily on the **experiment**. The experiment can also falsify MEPP (this allows considering MEPP as a physical principle). MEPP is to be tested using unambiguously interpreted, well-studied and fairly simple experimental systems (in this regard, at the present development stage of MEPP science, climatic, biological and similar systems are certainly not suitable for confirmation or falsification due to their complexity and ambiguous interpretation)[9]. Mathematical models are also absolutely unsuitable for falsification. So, a model is only some more or less crude and often one-sided reflection of some part of a phenomenon, whereas MEPP is the principle reflecting the dissipative properties that are observed in nature rather than in its model. We consider the investigation of nonequilibrium phase transitions in the homogeneous systems to be a possible way of MEPP falsification. For example, if the regime with the smallest entropy production (particularly, with the smallest generated heat of dissipation) proves to be the most probable (in the statistical sense) in the case of the fixed thermodynamic force (for example, the pressure or temperature gradient) and *several possible* existing nonequilibrium phases (regimes), then we can disprove MEPP (or at least considerably narrow the range of its applicability).

---

[8] Let us emphasize that MEPP states that the entropy production is not only above zero but is the maximum possible one.

[9] This will only bring unjustified discredit to the principle and result in a negative response thereto (see, e.g. [35, 36]).

Certainly, microscopic methods of principle justification are important for understanding the scope of MEPP validity. However, their significance in this case should not be exaggerated. The role of the statistical method (physics) has always been of rather supplementary, subordinate nature: scientists tried to understand and illustrate macroscopic properties that were experimentally discovered and integrally generalized in the thermodynamic postulates (laws) using simple models (of ideal gas, Markovian processes, etc.). If results of the statistical investigation were not proved experimentally and/or contradicted thermodynamics, the used statistical model was considered to be too crude or erroneous, and the model was modified (and the opposite has never occurred in physics). It is therefore methodologically incorrect to state that the microscopic view on the world and the entropy substitutes (analytically derives etc.) MEPP. It is as crude to state that as to claim, for example, that Boltzmann's H-theorem has become a proof of the second law of thermodynamics.

4) The existing interpretation of MEPP from the viewpoint of Jaynes' informational methodology should not be identified with the microscopic (statistical) interpretation. This is a special view on the foundation of the statistical physics, which has both its supporters and opponents (see e.g. [37]). The laconic brevity and the simplicity, which enables to formulate, for example, the foundation of the equilibrium statistical physics, can be reckoned among its benefits. The subjectivism of Jaynes' method can be considered as a drawback. There are attempts to obtain MEPP (the principle of nonequilibrium physics) from the maximization of informational (trajectory) entropy [4-11]. This is a very interesting area of studies. However, we would like to raise a number of criticisms:

♦ MEPP has still not been obtained in a sufficiently rigorous manner from the maximization of informational (trajectory) entropy [4-11]. Even if it is achieved, this will be done only with the involvement of **additional** important suppositions/assumptions. It is these suppositions that will prevent from concluding that MEPP is just a consequence; instead, they will indicate that MEPP is an independent statement.

♦ The informational approach is a variety of microscopic approach. The incompleteness of statistical methods for justifying the statements (principles) based on experiments has already been mentioned in item 3. These can also be repeated for the considered approach. However, it also has particular disadvantages. Thus, Jaynes' method is a procedure (often an effective one) of searching for the relations in the case of *known or supposed* constraints. If the result disagrees with the reality, then the constraints are replaced (or the priorities of the constraints are changed). The procedure itself (explicit form and properties of the informational entropy, existence and uniqueness of the solution, and the like) is postulated. As a result, this approach is useful as a simple *algorithm* of obtaining the known (generally accepted) solution[10]. If the solution is, however, unknown (or there are different opinions about the solution), then any sufficiently experienced scientist can use this method to obtain any desired result depending on his/her preferences[11]. In this regard, the use of a procedure (such as the informational one) rather than a physical law (objective by definition) for justifying MEPP (which is at the stage of its final formulation and understanding so far) provokes our objections. In other words, Jaynes' mathematical procedure can be used for obtaining, depending on the selected physical and mathematical constraints, multiple other procedures (the value of such mathematical exercises becomes, nevertheless, rather doubtful for physics); however, if the maximum entropy production is considered to be a physical principle, then it will be fundamentally unachievable using Jaynes' inference algorithm because MEPP itself is the *key physical constraint* (like the first/second law of thermodynamics or the charge conservation law).

---

[10] Indeed, the researcher's intention to mathematically make the most unprejudiced prediction in the conditions of incomplete information about the system is the essence of this method. Therefore, if a phenomenon is very poorly experimentally studied (there are not enough constraints), then anything can be predicted using the informational method (there are no truth criteria). As opposed to that, when MEPP or the maximum of entropy are considered as the physical laws/principles, the scientists have substantially less possibilities for "ungrounded fancies". It is the methods able to predict something specific for poorly studied phenomena (wherefrom their falsifiability is implied) that are especially valuable for natural science.

[11] If this cannot be achieved by selecting the constraints, then other kinds of informational entropy can always be used, for example, by Tsallis, Abe, Kullback, and many, many others [23, 30].

Figure Capture

Fig1. An example of some transitions of the system from one state to another. For equilibrium state $\boldsymbol{\alpha} = \boldsymbol{0}$. $P(\boldsymbol{\alpha}|\boldsymbol{\alpha_0}, \tau) < P(\boldsymbol{\alpha}|\boldsymbol{\alpha}^*(\boldsymbol{\alpha}), \tau) = \Xi = P(\boldsymbol{\alpha_0}|\boldsymbol{\alpha}^*(\boldsymbol{\alpha_0}), \tau) > P(\boldsymbol{\alpha_0}|\boldsymbol{\alpha}, \tau)$.

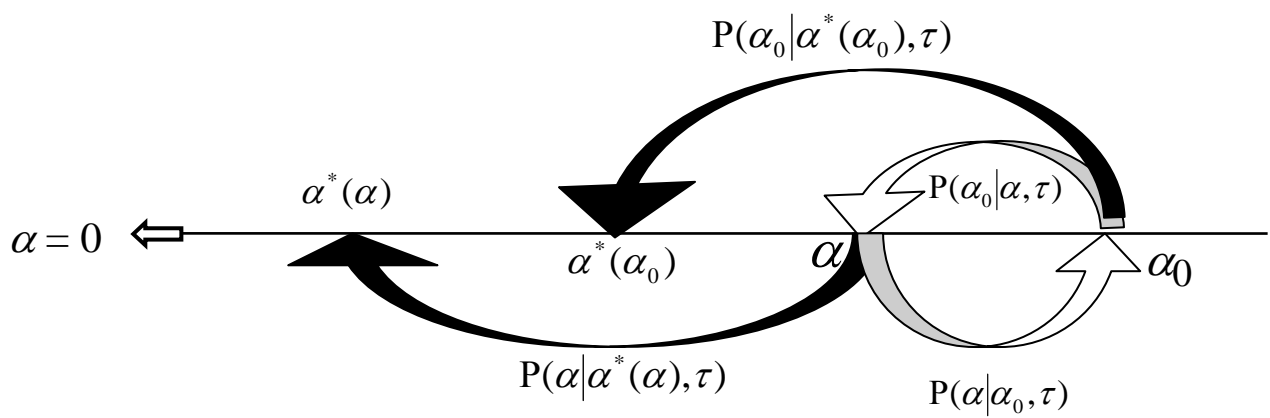